\documentclass{amsart}
\usepackage{amsmath}
\usepackage{graphicx}
\usepackage{amsfonts}
\usepackage{amssymb}

\newtheorem{theorem}{Theorem}
\theoremstyle{plain}

\newtheorem{corollary}{Corollary}

\newtheorem{definition}{Definition}

\numberwithin{equation}{section}

\begin{document}
\title{Generalizations of the Brachistochrone Problem}
\author{John A. Gemmer$^{1}$}
\address{Department of Applied Mathematics\\
University of Arizona\\
Tucson, AZ 85721}
\email{jgemmer@math.arizona.edu}
\author{Michael Nolan}
\address{Department of Physics\\
Millersville University of Pennsylvania\\
Millersville, PA. 17551}
\email{michael.nolan@millersville.edu}
\author{Ron Umble}
\address{Department of Mathematics\\
Millersville University of Pennsylvania\\
Millersville, PA. 17551}
\email{ron.umble@millersville.edu}
\date{January 27, 2009}
\thanks{$^{1}$ This paper contains the results in the first author's
undergraduate thesis codirected by the second and third authors }

\begin{abstract}
Consider a frictionless surface $S$ in a gravitational field that need not
be uniform. Given two points $A$ and $B$ on $S$, what curve is traced out by
a particle that starts at $A$ and reaches $B$ in the shortest time? This
paper considers this problem on simple surfaces such as surfaces of
revolution and solves the problem two ways: First, we use conservation of mechanical energy and the Euler-Lagange equation; second, we use geometrical optics and the eikonal equation.  We conclude with a discussion of the relativistic effects at relativistic velocities. 

\end{abstract}

\maketitle

\section{Introduction}

In 1696, Johann Bernouli  posed the following \textquotedblleft
Brachistochrone problem\textquotedblright\: \textit{Find the shape of the
curve down which a bead sliding from rest and accelerated by gravity will
fall from one point to another in the least time.} This problem assumes that
the particle is falling on a vertical plane in a uniform gravitational
field. Sir Isaac Newton, Gottfried Leibniz, Guillaume De L'Hospital, Jakob Bernouli, and Johann Bernouli showed that the solution is a cycloid, the curve
traced out by a point on the rim of a rolling circle, see Dunham \cite{Dunham}. Solutions of the classical Brachistochrone problem typically use techniques
of calculus of variations, see Gelfand and Fomin \cite{Gelfand}, or geometrical optics, see Erlichson \cite{Erlichson}.

The problem of finding Brachistochrone curves with coulomb friction lying on a vertical plane in a uniform gravitational field has been discussed by Ashby et al. \cite{Ashby}, Hayen \cite{Hayen}, Heijden and Diepstraten \cite{Heijden}, and further generalized to a curve with friction lying on a cylinder by Covic and Veskovic \cite{Covic}. Vratanar and Saje \cite{Vratanar} discuss the related problem of finding the brachistochrone curve in a non-conservative resistance field. The problem of finding Brachistochrone curves on cylinders in uniform fields has been solved by Yamani and Mulhem \cite{Yamani} and on cylinders and spheres and in the unpublished work of Palmieri \cite{Palmieri}. The generalization to non-uniform fields has been discussed in the works of Aravind \cite{Aravind}, Denman \cite{Denman}, and Venezian \cite{Venezian}, who find Brachistochrone solutions in linear radial fields, and the works of Denman \cite{Denman}, Parnovsky \cite{Parnovsky}, and Tee \cite{Tee}, who find solutions in inverse square radial fields. Further generalizations of this problem to include special relativistic effects have been studied by Farina \cite{Farina}, Goldstein and Bender \cite{GoldsteinandBender}, and Scarpello and Ritelli \cite{Scarpello}.

This paper considers the following generalized problem: \textit{Let }$S$ \textit{be a smooth frictionless surface in a (not necessarily uniform) gravitational field. Given two points }$A$ \textit{and} $B$ \textit{on} $S$%
\textit{, find the curve from }$A$ \textit{to} $B$ \textit{along which a particle released from }$A$ \textit{reaches} $B$\textit{\ in minimal time.} In section 2 of this paper we use the Euler-Lagrange equation to present a theorem that solves the problem for a large class of surfaces in various gravitational fields. Our solution is general enough so that Yamani and Mulhem' \cite{Yamani} solution on a cylinder is a direct application of our theorem as is Parnovsky \cite{Parnovsky} and Tee's \cite{Tee} solution to the inverse square problem. Furthermore, generalizing the result of Parnovsky \cite{Parnovsky} and Tee \cite{Tee}, we prove that for a central force field with $V=-r^{-n}$ no solution curve enters the sector $-2\pi/(n+2)\leq \theta \leq 2\pi /(n+2)$. In section 3 we generalize the optico-mechanical analogy used by Aravind \cite{Aravind}, Farina \cite{Farina}, and Parnovsky \cite{Parnovsky} to surfaces. In particular, we use the eikonal equation \cite{Sommerfeld} to provide another proof of the theorem proved in section 2. 
Using the eikonal equation we find a solution for a charged particle moving at relativistic velocities lying in a uniform electric field which coincides with Goldstein and Bender's \cite{GoldsteinandBender} and Farina's \cite{Farina} result. Finally, we apply the light ray curvature equation \cite{Sommerfeld} to prove that torsion always vanishes along Brachistochrone solutions in a central force fields. This result proves the implicit or explicit assumption taken by Aravind \cite{Aravind}, Denman \cite{Denman}, Venezian \cite{Venezian}, Parnovsky \cite{Parnovsky}, and Tee \cite{Tee} that Brachistochrone solutions in radial fields are planar curves. 

The numerical computations and graphics in this paper were carried out and created in Mathematica \cite{Mathematica}.  

\section{Solutions using Calculus of Variations}

Let $S$ be a smooth frictionless surface in some gravitational field and let 
$V\left( P\right) $ be the gravitational potential at point $P$ on $S$.
Choose distinct points $A$ and $B$ with $V(A)>V(B)$ and assume that the
velocity of a particle falling along a smooth curve from $A$ to $B$ is much
less than the speed of light. Then by Newtonian conservation of mechanical
energy we have 
\begin{equation*}
\frac{1}{2}\left( \frac{ds}{dt}\right) ^{2}+V=V(A),
\end{equation*}
where $ds$ is the arclength on $S$. Solving for $\left( \frac{ds}{dt}\right)
^{2}$ gives 
\begin{equation*}
2(V(A)-V)=\left( \frac{ds}{dt}\right) ^{2}
\end{equation*}%
and separating variables yields%
\begin{equation*}
dt=\pm \sqrt{\frac{1}{2(V(A)-V)}}\,ds.
\end{equation*}%
Thus the total time $T$ is given by 
\begin{equation}
T=\pm \sqrt{\frac{1}{2}}\int_{A}^{B}\sqrt{\frac{1}{V(A)-V}}\,ds.
\label{time}
\end{equation}

Now, let $U$ be an open subset of $\mathbb{R}^{3}$ and let $\mathbf{x}%
:U\rightarrow S$ be an orthogonal coordinate patch. Then $\mathbf{x}\left(
u,v\right) =\left( x_{1}(u,v),x_{2}(u,v),x_{3}(u,v)\right) $ is a subset of 
$S$ in which $ds^{2}=Edu^{2}+Gdv^{2}$, where $E$ and $G$ are the metric
coefficents $E=\frac{\partial \mathbf{x}}{\partial u}\bullet \frac{\partial 
\mathbf{x}}{\partial u}$ and $G=\frac{\partial \mathbf{x}}{\partial v}%
\bullet \frac{\partial \mathbf{x}}{\partial v}$, and we may rewrite equation
(\ref{time}) in the form 
\begin{equation}
T=\pm \sqrt{\frac{1}{2}}\int_{A}^{B}\sqrt{\frac{1}{V(A)-V}}\,\sqrt{%
Edu^{2}+Gdv^{2}}  \label{time!}
\end{equation}

\noindent and either as\ 
\begin{equation}
T=\pm \sqrt{\frac{1}{2}}\int_{A}^{B}\sqrt{\frac{E\left( \frac{du}{dv}\right)
^{2}+G}{V(A)-V}}\,dv=\pm \sqrt{\frac{1}{2}}\int_{A}^{B}\mathcal{F}%
(u,v,u^{\prime })dv  \label{time1}
\end{equation}%
or%
\begin{equation}
T=\pm \sqrt{\frac{1}{2}}\int_{A}^{B}\sqrt{\frac{E+G\left( \frac{dv}{du}%
\right) ^{2}}{V(A)-V}}\,du=\pm \sqrt{\frac{1}{2}}\int_{A}^{B}\mathcal{F}%
(u,v,v^{\prime })du,  \label{time2}
\end{equation}

\noindent where $^{``\prime }\textquotedblright $ denotes differentiation
with respect to the other variable.

To minimize $T$ in (\ref{time1}), we must solve the Euler-Lagrange equation\ 
\begin{equation}
\frac{d}{dv}\frac{\partial \mathcal{F}}{\partial u^{\prime }}-\frac{\partial 
\mathcal{F}}{\partial u}=0.  \label{euler}
\end{equation}%
Since $\frac{\partial E}{\partial u}=\frac{\partial G}{\partial u}=\frac{%
\partial V}{\partial u}=0$ implies $\frac{\partial \mathcal{F}}{\partial u}%
=0 $ (see \cite{Gelfand}), equation (\ref{euler}) simplifies to 
\begin{equation*}
\frac{\partial \mathcal{F}}{\partial u^{\prime }}=C,
\end{equation*}%
where $C$ is a constant that depends upon $B$. By differentiating $\mathcal{F%
}$ with respect to $u^{\prime }$ we obtain the separable differential
equation 
\begin{equation*}
\pm \frac{Eu^{\prime }}{\sqrt{(V(A)-V)(E(u^{\prime })^{2}+G)}}=C.
\end{equation*}%
Squaring both sides and simplifying gives 
\begin{equation*}
\frac{du}{dv}=\pm \sqrt{\frac{C^{2}(V(A)-V)G}{E(E-C^{2}(V(A)-V)))}}
\end{equation*}%
whose solution expresses $u$ as a function of $v$ (see item 1 in Theorem \ref%
{one} below). Similarly, since $\frac{\partial \mathcal{F}}{\partial v}=0$
when $\frac{\partial E}{\partial v}=\frac{\partial G}{\partial v}=\frac{%
\partial V}{\partial v}=0,$ we may use equation (\ref{time2}) and a modified
form of (\ref{euler}) to obtain $v$ as a function of $u.$

\begin{theorem}
\label{one}Let $\mathbf{x}:U\rightarrow S$ be an orthogonal coordinate patch
on a smooth frictionless surface $S$.

\begin{enumerate}
\item If $\frac{\partial E}{\partial u}=\frac{\partial G}{\partial u}=\frac{%
\partial V}{\partial u}=0$, then the solution to the Brachistochrone problem
on $S$ is given by the curve $\mathbf{x}(u(v),v)$, where 
\begin{equation}
u\left( v\right) =\pm \int_{A}^{v}\sqrt{\frac{C^{2}G(w)(V(A)-V(w))}{%
E(w)[E(w)-C^{2}(V(A)-V(w))]}}\,dw.  \label{solution1}
\end{equation}%
\ 

\item If $\frac{\partial E}{\partial v}=\frac{\partial G}{\partial v}=\frac{%
\partial V}{\partial v}=0$, then the solution to the Brachistochrone problem
on $S$ is given by the curve $\mathbf{x}(u,v(u))$, where%
\begin{equation}
v\left( u\right) =\pm \int_{A}^{u}\sqrt{\frac{C^{2}E(w)(V(A)-V(w))}{%
G(w)[G(w)-C^{2}(V(A)-V(w))]}}\,dw.  \label{solution2}
\end{equation}
\end{enumerate}
\end{theorem}

\subsection{Applications of Theorem \protect\ref{one}}

\subsubsection{The Classical Brachistochrone Problem}

Let us apply Theorem \ref{one} to solve the classical Brachistochrone
Problem. In this case, the particle is falling on the vertical plane given
by $\mathbf{x}(u,v)=(u,0,v).$ Since $E=G=1$ and $F=0,$ we obtain 
\begin{equation*}
u(v)=\pm \int_{a}^{v}\sqrt{\frac{C^{2}(a-v)}{1-C^{2}(a-v)}}\,dw,
\end{equation*}%
where $V(v)=v$ and $a$ is the particle's initial $v$-coordinate. Figure \ref%
{cycloids} illustrates several solution curves for a particle starting at
the origin, each of which is uniquely determined by $C$ and a choice of
sign. Physically, the value of $C$ is determined by the destination point $B$
and the sign indicates motion to the left or right. Note that each of these
curves represents a family of solution curves since the terminal point $B$
can be positioned anywhere along the curve.

\begin{figure}[ht]
\begin{center}
\includegraphics[width=4.0in]{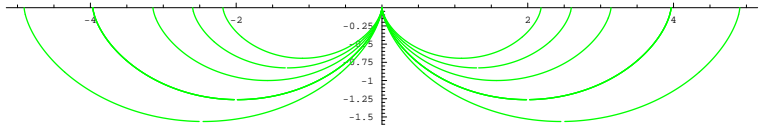}
\end{center}
\caption{Cycloid solutions in a uniform field}
\label{cycloids}
\end{figure}

\subsubsection{Surfaces of Revolution}

Let us consider the Brachistochrone problem on smooth surfaces of revolution 
$S$ parametrized by the coordinate patches 
\begin{eqnarray*}
\mathbf{x}_{n}(u,v) &=&(h(u)\cos v,h(u)\sin v,g(u)),\text{ }\left(
u,v\right) \in \mathbb{R\times }\left( (2n-1)\pi ,(2n+1)\pi \right)  \\
\mathbf{y}_{n}(u,v) &=&(h(u)\cos v,h(u)\sin v,g(u)),\text{ }\left(
u,v\right) \in \mathbb{R\times }\left( 2n\pi ,(2n+2)\pi \right) ,
\end{eqnarray*}%
where $g,h:\mathbb{R\rightarrow R}$ are differentiable
functions with $h>0$, $n\in \mathbb{Z}$, and 
\begin{equation*}
\mathbf{y}_{n}\left( u,v\right) =\left\{ 
\begin{array}{ll}
\mathbf{x}_{n}\left( u,v\right) , & v\in \left( 2n\pi ,(2n+1)\pi \right)  \\ 
\mathbf{x}_{n+1}\left( u,v\right) , & v\in \left( (2n+1)\pi ,(2n+2)\pi
\right) .%
\end{array}%
\right. 
\end{equation*}

\begin{corollary}
\label{cor1}If $S$ is in a uniform gravitational field parallel to the $z$%
-axis, the solution to the Brachistochrone Problem on $S$ for a particle 
with initial position $A=(h(u_{0}),0,g(u_{0}))$ is
the curve 
\begin{equation*}
\gamma \left( u\right) =\left\{ 
\begin{array}{ll}
\mathbf{x}_{n}\left( u,v\left( u\right) \right) , & v\left( u\right) \in 
\left[ 2n\pi ,(2n+1)\pi \right)  \\ 
\mathbf{y}_{n}\left( u,v\left( u\right) \right) , & v\left( u\right) \in 
\left[ (2n+1)\pi ,(2n+2)\pi \right) ,%
\end{array}%
\right. 
\end{equation*}%
where\ 
\begin{equation*}
v(u)=\pm \int_{u_{0}}^{u}\sqrt{\frac{C^{2}(h^{\prime }(w)^{2}+g^{\prime
}(w)^{2})(g(u_{0})-g(w))}{h(w)^{2}((h(w)^{2}-C^{2}(g(u_{0})-g(w)))}}\,dw.
\end{equation*}
\end{corollary}

\begin{figure}[th]
\centerline{
    \mbox{\includegraphics[height=2.0in]{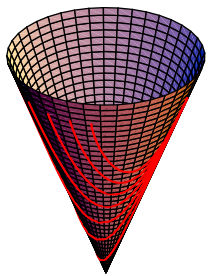}}
  }
\caption{Solution curves on the cone}
\label{cone}
\end{figure}

The right circular cone is obtained as a surface of revolution $S$ by setting $h\left(
u\right) =g(u)=u>0$. For a particle with initial position $A=(1,0,1)$, 
the solution
curves $\gamma \left( u\right) $ on $S$ given by Corollary \ref{cor1} 
are defined in terms of 
\begin{equation*}
v\left( u\right) =\pm \int_{u_{0}}^{u}\sqrt{\frac{2C^{2}(u_{0}-w)}{%
w^{2}(w^{2}-C^{2}(u_{0}-w))}}\,dw,
\end{equation*}%
where the sign $\pm $ positions the curve in one of the two half-spaces
determined by the $yz$-plane. Figure \ref{cone} illustrates several solution
curves.

Another interesting surface of revolution is the hyperboloid of one 
sheet obtained by setting $h\left( u\right) =\cosh u$
and $g\left( u\right) =\sinh u$. For a particle with initial position $A=(1,0,0)$,
the solution curves $\gamma \left( u\right) $  given by Corollary \ref{cor1} are defined
in terms of  
\begin{equation*}
v\left( u\right) =\pm \int_{u_{0}}^{u}\sqrt{\frac{C^{2}(\sinh u_{0}-\sinh
w)((\cosh w)^{2}+(\sinh w)^{2}))}{(\cosh w)^{2}((\cosh w)^{2}-C^{2}(\sinh
u_{0}-\sinh w))}}\,dw.
\end{equation*}%
But unlike solution curves on the right circular cone, some solution curves on the
hyperboloid do not attain a minimum $z$-coordinate but continue downward,
spiraling around the hyperboloid and intersecting other solution curves as
they do. When two spiraling solution curves intersect, one must evaluate the
integral in (\ref{time!}) to determine which curve minimizes time.

\begin{figure}[ht]
\centerline{
    \mbox{\includegraphics[height=2.5in]{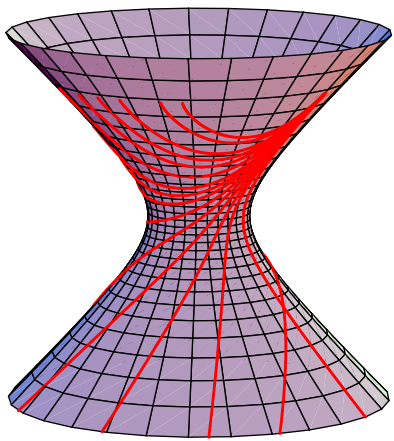}}
  }
\caption{Solution curves on the hyperboloid of one-sheet}
\label{hyperboloid}
\end{figure}

\subsubsection{Central Force Fields}

\begin{figure}[ht]
\centerline{
    \mbox{\includegraphics[width=2.5in]{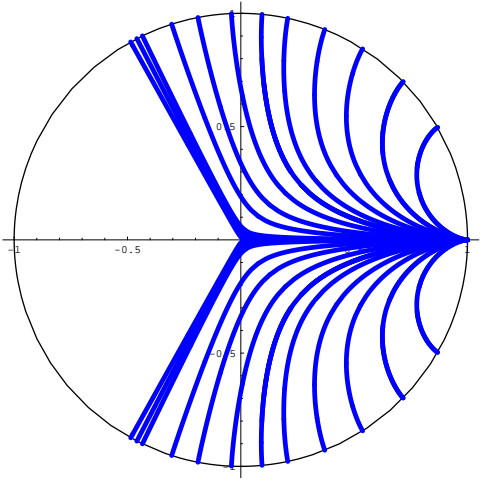}}
  }
\caption{Solution curves for the inverse square field ($n=1$).}
\label{inverse}
\end{figure}

Consider a particle confined to a plane and falling in a central force field
with $V=-\frac{1}{u^{n}}$ for some $n>0$. Using polar coordinates, we can
parametrize $\mathbb{R}^{2}$ by $\mathbf{x}(u,v)=(u\cos v,u\sin v),$ $\left(
u,v\right) \in \left( 0,+\infty \right) \times \left( -\pi,\pi \right)$; then the metric coefficients are simply $E=1\,,F=0\,$ and $G=u^{2}$. Now apply Theorem 1 and set $D=1/C^{2}$ to obtain the solution curves 
\begin{eqnarray}
v(u) &=&\pm \int_{u_{0}}^{u}\sqrt{\frac{C^{2}(-\frac{1}{u_{0}^{n}}+\frac{1}{%
w^{n}})}{w^{2}[w^{2}-C^{2}(-\frac{1}{u_{0}^{n}}+\frac{1}{w^{n}})]}}dw
\label{inverse2} \\
&&  \notag \\
&=&\pm \int_{u_{0}}^{u}\sqrt{\frac{(u_{0}^{n}-w^{n})}{
w^{2}(w^{n+2}u_{0}D-(u_{0}^{n}-w^{n}))}}\,dw,  \notag
\end{eqnarray}
where $A=\mathbf{x}(u_{0},0)$ is the initial position of $A$. Figure 
\ref{inverse} illustrates several solution curves for an inverse square
field with $n=1$. In this case, the $\pm $ sign positions $A$ in the first
or fourth quadrant.

Interestingly, when $-\pi <v<0,$ numerical plots of solution curves
asymptotically approach the line $\theta =-2\pi /3$ as $C\rightarrow 0.$
Since solution curves in the region $0<v<\pi $ are mirror images of those in
the region $-\pi <v<0$, these calculations suggest that no solution curve
enters the sector $-2\pi /3\leq \theta \leq 2\pi /3$. Note that solution
curves seem to approach one of two limit curves that follow the $x$-axis to
the origin then continue along the rays $\theta =-2\pi /3$ or $\theta =2\pi
/3 $. This is indeed the case and was observed and proved by Parnovsky \cite{Parnovsky} and Tee \cite{Tee}. The following is an extension of their work for all $n>0$.

Without loss of generality assume that $A=\mathbf{x}(1,0)$ and the sign in (%
\ref{inverse2}) is positive so that the particle's trajectory lies in the fourth
quadrant. Note that when 
\begin{equation}
\frac{du}{dv}=0  \label{critequation}
\end{equation}%
the particle begins to move away from the origin and the sign in (\ref%
{inverse2}) is reversed. We will show that equation (\ref{critequation}) holds for exactly one positive real value by investigating the zeros of the polynomial 
\begin{equation}
r(w)=w^{n+2}D+w^{n}-1.  \label{turningpoint}
\end{equation}%
Since $r(0)=-1$ and $r(1)=D$, the Intermediate Value Theorem tells us that $r(c_0)=0$ for some $c_0 \in (-1,1)$.
Furthermore, 
\begin{equation*}
r^{\prime }(w) =(n+2)w^{n+1}D+nw^{n-1}=w^{n-1}\left( (n+2)w^{2}D+n\right) .
\end{equation*}%
But $r^{\prime }(w)>0$ for all $0<w<D$ so $r$ is increasing and $c_{0}$ is
unique. Let 
\begin{equation}
\theta =\int_{1}^{c_{0}}\sqrt{\frac{(1-w^{n})}{w^{2}(w^{n+2}D-(1-w^{n}))}}%
\,dw.  \label{theta}
\end{equation}%
Since solution curves are symmetric in the line through $\mathbf{x}%
(c_{0},\theta )$ and the origin, we can determine the least upper bound of the polar angle $\theta$  by evaluating its limit as $c_{0}\rightarrow 0$. Using
equation (\ref{turningpoint}) we can express $D$ in terms of $c_{0}$ by 
\begin{equation*}
D=c_{0}^{-n-2}-c_{0}^{-2}.
\end{equation*}%
Then, equation (\ref{theta}) can be rewritten as 
\begin{eqnarray}
\theta  &=&\int_{1}^{c_{0}}\sqrt{\frac{(1-w^{n})}{%
w^{2}(w^{n+2}(c_{0}^{-n-2}-c_{0}^{-2})-(1-w^{n}))}}\,dw  \label{integral} \\
&&  \notag \\
&=&c_{0}^{\frac{n+2}{2}}\int_{1}^{c_{0}}\frac{1}{w}\sqrt{\frac{1-w^{n}}{%
w^{n+2}(1-c_{0}^{n})-c_{0}^{n+2}(1-w^{n})}}\,dw  \notag
\end{eqnarray}%
Let $x=\left( \frac{c_{0}}{w}\right) ^{\frac{n+2}{2}}$; then (\ref{integral}%
) reduces to 
\begin{eqnarray*}
\theta  &=&-\frac{2}{n+2}c_{0}^{\frac{n+2}{2}}\int_{c_{0}^{\frac{n+2}{2}%
}}^{1}\frac{1}{x}\sqrt{\frac{1-c_0^nx^{\frac{-2n}{n+2}}}{%
c_{0}^{n+2}x^{-2}(1-c_{0}^{n})-c_{0}^{n+2}\left(1-c_{0}^{n}x^{\frac{-2n}{n+2}}\right)}%
}\,dx \\
&& \\
&=&-\frac{2}{n+2}\int_{c_{0}^{\frac{n+2}{2}}}^{1}\sqrt{\frac{1-c_{0}x^{\frac{%
-2n}{n+2}}}{(1-c_{0}^{n})-x^{2}\left(1-c_{0}^{n}x^{-\frac{2n}{n+2}}\right)}}\,dx.\\
&&\\
&=& -\frac{2}{n+2}\int_{c_0^{\frac{n+2}{2}}}^{1}\sqrt{\frac{1}{\frac{1-c_0^n}{1-c_0^nx^{-\frac{2n}{n+2}}}-x^2}}\,dx
\end{eqnarray*}%
Now, for $c_0^{\frac{n+2}{n}}<x<1$, we have that $\frac{1-c_0^n}{1-c_0^nx^{-\frac{2n}{n+2}}}>1$. Therefore,
\begin{equation*}
\theta=-\frac{2}{n+2}\int_{c_0^{\frac{n+2}{2}}}^{1}\sqrt{\frac{1}{\frac{1-c_0^n}{1-c_0^nx^{-\frac{2n}{n+2}}}-x^2}}\,dx\leq \int_0^1 \sqrt{\frac{1}{1-x^2}}\,dx.
\end{equation*}
\noindent Finally, it follows by an application of the Dominated Convergence Theorem that  
\begin{equation*}
\lim_{c_{0}\rightarrow 0}\theta =-\frac{2}{n+2}\int_{0}^{1}\frac{1}{\sqrt{1-x^{2}}}\,dx=-\frac{2}{n+2}\left( \arcsin (1)-\arcsin (0)\right)=-\frac{\pi }{(n+2)}.
\end{equation*}

Now, by symmetry the maximum polar angle of the particle is less than twice this limiting  value. Therefore the limiting polar angle of a particle falling on a solution curve is $
\frac{2\pi }{n+2}$ and the central angle of the \textquotedblleft forbidden
sector\textquotedblright\ is $2\pi -\frac{4\pi }{n+2}=\frac{2n\pi }{n+2}$. \
In summary, we have proved:

\begin{theorem}
\label{thm2}Let $A\left( r,\phi \right) \in \mathbb{R}^{2},$ $r>0.$ If $V=-%
\frac{1}{u^{n}},$ $n>0,$ every Brachistochrone curve initiating at $A$
lies in the sector $\phi -\frac{2\pi }{n+2}\leq \theta \leq \phi +\frac{2\pi 
}{n+2}$.
\end{theorem}

\section{Solutions using Geometrical Optics}

In this section we approach the Brachistochrone problem from the perspective
of geometrical optics by considering the path of a light ray in a medium of
nonuniform index of refraction. Let $S:\mathbf{x}\left( u,v\right) $ be a
smooth surface in such a medium with index of refraction $n\left( u,v\right)
$. According to Fermat's Principle, light propagated from source $A$
minimizes the total time $T$ of travel to destination $B$, i.e., 
\begin{equation}
T=\frac{1}{c}\int_{A}^{B}n(u,v)\,ds,  \label{optical length}
\end{equation}%
where $c$ is the speed of light in a vacuum. Note that when $n(u,v)=\sqrt{%
\frac{1}{V(A)-V(u,v)}}$ and $c=1\,$m/sec, equation (\ref{optical length}) is
equivalent to equation (\ref{time}). Therefore, computing the path of a
light ray in this medium simultaneously produces a Brachistochrone solution.

\subsection{The Eikonal Equation}

One can describe light rays in terms of their wavefronts, which are level
surfaces of some differentiable function $L:\mathbb{R}^{3}\rightarrow 
\mathbb{R}$. More precisely, given a light ray $\alpha $ passing through a
medium with index of refraction $n$, let $\mathbf{T}$ be the unit tangent.
Then $L$ is determined by the eikonal equation 
\begin{equation}
\nabla L=n\mathbf{T}.  \label{eikonal}
\end{equation}%
We can use (\ref{eikonal}) to solve the Brachistochrone Problem without
appealing to the Euler-Lagrange equation. Assume that $\alpha $ is
constrained to some smooth surface $S$ with metric $ds^{2}=Edu^{2}+Gdv^{2}$
in some medium with index of refraction $n(u,v)$. Suppose $\frac{\partial E}{%
\partial v}=\frac{\partial G}{\partial v}=\frac{\partial n}{\partial v}=0$
and that  
\begin{equation}
L=Cv+f(u).  \label{eikonalcurve}
\end{equation}%
Then equation (\ref{eikonal}) yields the differential equation 
\begin{equation*}
\nabla L\bullet \nabla L=G^{-1}C^{2}+E^{-1}\left( \frac{df}{du}\right)
^{2}=n^{2}.
\end{equation*}%
Solving this separable equation for $f$ gives%
\begin{equation}
f(u) =\pm \int \sqrt{E\left( n^{2}-C^{2}/G\right) }\,du=\pm \int \sqrt{\frac{E(G-C^{2}/n^{2})}{G/n^{2}}}\,du. \label{f(u)}
\end{equation}
Using equations (\ref{eikonal}) and (\ref{eikonalcurve}), and letting $%
\mathbf{e}_{u}=\frac{\mathbf{x}_{u}}{\Vert \mathbf{x}_{u}\Vert }$ and $%
\mathbf{e}_{v}=\frac{\mathbf{x}_{v}}{\Vert \mathbf{x}_{v}\Vert }$ we have 
\begin{equation*}
n\mathbf{T}=\sqrt{\frac{G-C^{2}/n^{2}}{G/n^{2}}}\mathbf{e}%
_{u}+CG^{-\frac{1}{2}}\mathbf{e}_{v}.
\end{equation*}%
Expressing $\mathbf{T}$\textbf{\ }in differential form gives 
\begin{equation*}
n\left( E^{\frac{1}{2}}\frac{du}{ds}\mathbf{e}_{u}+G^{\frac{1}{2}}\frac{dv}{%
ds}\mathbf{e}_{v}\right) =\sqrt{\frac{G-C^{2}/n^{2}}{G/n^{2}}}%
\mathbf{e}_{u}+CG^{-\frac{1}{2}}\mathbf{e}_{v}.
\end{equation*}%
Therefore 
\begin{equation*}
\frac{dv}{du}\frac{G^{\frac{1}{2}}}{E^{\frac{1}{2}}}=\sqrt{\frac{C^{2}/n^{2}%
}{G-C^{2}/n^{2}}}.
\end{equation*}%
 By solving this equation we obtain:

\begin{theorem}
\label{thm3}Let $\mathbf{x}:U\rightarrow S$ be an orthogonal coordinate
patch on a smooth frictionless surface $S$.

\begin{enumerate}
\item If $\frac{\partial E}{\partial u}=\frac{\partial G}{\partial u}=\frac{%
\partial n}{\partial u}=0$, the light rays on $S$ are given by $\mathbf{x}%
(u(v),v)$, where\ 
\begin{equation*}
u(v)=\pm \int_{v_{0}}^{v}\sqrt{\frac{GC^{2}/n^{2}}{E(E-C^{2}/n^{2})}}\,dw.
\end{equation*}

\item If $\frac{\partial E}{\partial v}=\frac{\partial G}{\partial v}=\frac{%
\partial n}{\partial v}=0$, the light rays on $S$ are given by $\mathbf{x}%
(u,v(u))$, where 
\begin{equation*}
v(u)=\pm \int_{u_{0}}^{u}\sqrt{\frac{EC^{2}/n^{2}}{G(G-C^{2}/n^{2})}}\,dw.
\end{equation*}
\end{enumerate}
\end{theorem}

\begin{corollary}
Let $\mathbf{x}:U\rightarrow S$ be an orthogonal coordinate patch on a
smooth frictionless surface $S$ in a medium with index of refraction $n^{2}=%
\frac{1}{V_{0}-V}$.

\begin{enumerate}
\item If $\frac{\partial E}{\partial u}=\frac{\partial G}{\partial u}=\frac{%
\partial V}{\partial u}=0$, we recover Theroem 1.1.

\item If $\frac{\partial E}{\partial v}=\frac{\partial G}{\partial v}=\frac{%
\partial V}{\partial v}=0$, we recover Theorem 1.2.
\end{enumerate}
\end{corollary}

\subsection{Special Relativistic Solutions}

The strength of Theorem \ref{thm3} lies in its ability to give
Brachistochrone solutions in situations where classical conservation of
mechanical energy does not hold. For example, if the particle's velocity is
near the speed of light, to minimize the time of travel in the lab frame (see \cite{Goldstein}) we must replace the Newtonian mechanical energy
equation by its special relativistic counterpart 
\begin{equation}
\gamma c^{2}-c^{2}+V=V_{0},  \label{relenergy}
\end{equation}

\noindent where $c$ is the speed of light in a vacuum, $V$ is the gravitational potential, and $\gamma =(1-(\frac{ds}{dt})^{2}c^{-2})^{-\frac{1%
}{2}}$ (see \cite{Goldstein}). Solving (\ref{relenergy}) for $\gamma $ and
then for $(\frac{ds}{dt})^{2}$ gives 
\begin{equation*}
\left( \frac{ds}{dt}\right) ^{2}=\frac{c^{2}(V_{0}-V)(V_{0}-V+2c^{2})}{%
(V_{0}-V+c^{2})^{2}}.
\end{equation*}%
From equation (\ref{optical length}) it follows that 
\begin{equation}
\left( \frac{1}{n}\right) ^{2}=\frac{(V_{0}-V)(V_{0}-V+2c^{2})}{%
(V_{0}-V+c^{2})^{2}}.
\end{equation}

\noindent So, in sufficiently nice geometrical settings with reasonable
potentials, we can apply Theorem \ref{thm3} to find relativistic
Brachistochrone solutions.

\begin{figure}[ht]
\centerline{
    \mbox{\includegraphics[width=5.0in]{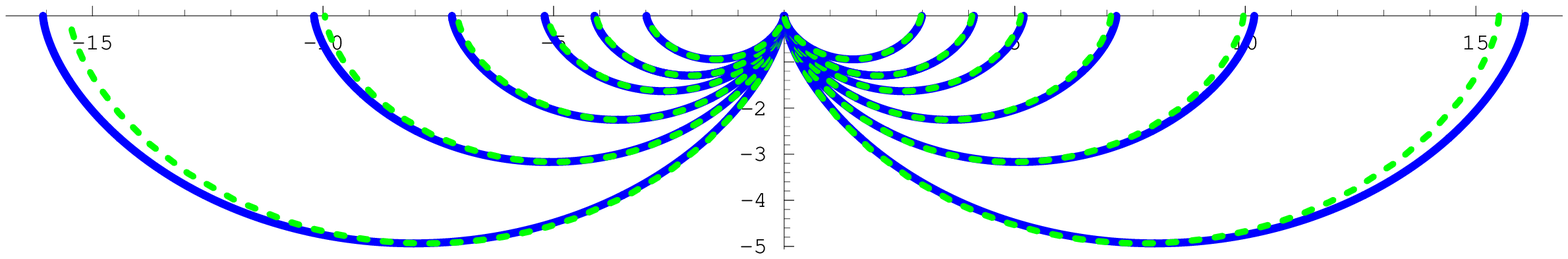}}
  }
\caption{Relativistic and classical Brachistochrone solutions.}
\label{Relativistica}
\end{figure}

\begin{figure}[ht]
\centerline{
    \mbox{\includegraphics[width=5.2in]{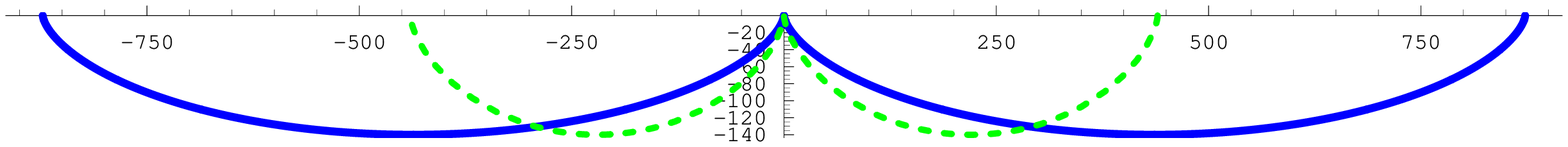}}
  }
\caption{Relativistic and classical Brachistochrone solutions.}
\label{Relativisticb}
\end{figure}

For example, let us again consider a particle falling in a uniform
gravitational field but confined to the vertical plane $\mathbf{x}%
(u,v)=(u,0,v)$. If the particle is falling at relativistic speeds we have 
\begin{equation*}
\left( \frac{1}{n}\right) ^{2}=\frac{(v_{0}-v)(v_{0}-v+2c^{2})}{%
(v_{0}-v+c^{2})^{2}},
\end{equation*}%
where $v_{0}$ is the initial height of the particle. Note, in actuality the gravitational potential energy depends upon the relativistic mass and not the rest mass. Therefore, the potential we have chosen corresponds to a charged particle under the influence of a uniform electric field (see Goldstein \cite{Goldstein} and Goldstein and Bender \cite{GoldsteinandBender}).  Setting $v_{0}=0$ and
applying Theorem \ref{thm3} gives the solution curves $\mathbf{x}(u(v),v)$,
where 
\begin{eqnarray*}
u(v) &=&\pm \int_{0}^{v}\sqrt{\frac{k^{2}/n^{2}}{1-k^{2}/n^{2}}}\,dw \\
&& \\
&=&\pm \int_{0}^{v}\sqrt{-\frac{k^{2}(2c^{2}-w)w}{%
c^{4}+2c^{2}(k^{2}-1)w-(k^{2}-1)w^{2}}}\,dw
\end{eqnarray*}%
and to avoid confusion we have indicated the constant of integration by $k$.
Figure \ref{Relativistica} illustrates several relativistic solution curves
and classical Brachistochrone solutions with $c=10\,$m/sec. The relativistic
solutions are plotted in solid blue; the classical solutions are plotted
with dashed green lines and closely approximate relativistic solutions
in this region. However, Figure \ref{Relativisticb} illustrates situations
in which such approximations are poor.

\subsection{Curvature of Light Rays}

Finally, we consider the curvature of light rays in a medium with index of
refraction $n$. Recall the following definitions and theorem from the
classical theory of curves (see \cite{McCleary}):

\begin{definition}
Let $\alpha (s)$ be a light ray with arc length parameter $s$ and let $%
\mathbf{\mathbf{T}}\left( s\right) $ be the unit tangent vector field along $%
\alpha $. The \underline{curvature} of $\alpha (s)$ is defined by $\kappa
=\left\Vert \frac{d\mathbf{\mathbf{T}}}{ds}\right\Vert $. If $\kappa \not=0$%
, the \underline{unit normal and binormal vector fields} along $\alpha $ are
given by $\mathbf{\mathbf{N}}=\frac{1}{\kappa }\mathbf{\mathbf{T}}^{\prime }$
and $\mathbf{\mathbf{B}}=\mathbf{\mathbf{T}}\times \mathbf{\mathbf{N}},$
respectively .
\end{definition}

\begin{theorem}
\label{thm5}\textbf{(Frenet-Serret). }If $\alpha (s)$ is a curve with arc
length parameter $s$ such that $\kappa (s)\neq 0$ for all $s$, then%
\begin{equation*}
\begin{array}{l}
\mathbf{\mathbf{T}}^{\prime }=\kappa \mathbf{\mathbf{N}}\text{, \ }\mathbf{%
\mathbf{N}}^{\prime }=-\kappa \mathbf{\mathbf{T}}(s)+\tau \mathbf{\mathbf{B}}%
\text{ \ and \ }\mathbf{\mathbf{B}}^{\prime }=-\tau \mathbf{\mathbf{N}}.%
\end{array}%
\end{equation*}
\end{theorem}

\begin{definition}
The function $\tau (s)$ is called the \underline{torsion} of $\alpha (s)$.
\end{definition}

Now, light rays are distinguished curves with the following property: 
\begin{equation}
\nabla n=n\kappa \mathbf{\mathbf{N}}+(\mathbf{\mathbf{T}}\bullet \nabla n)%
\mathbf{\mathbf{T}}.  \label{light ray curvature}
\end{equation}

In fact, when falling in a central force field, the torsion vanishes. To see
this, differentiate (\ref{light ray curvature}) and get 
\begin{eqnarray*}
\mathbf{\mathbf{N}}^{\prime } &=&\left( \frac{1}{\kappa n}\right) ^{\prime
}\kappa n\mathbf{\mathbf{N}}+\frac{1}{\kappa n}\left[ (\nabla n)^{\prime
}-\kappa (\mathbf{\mathbf{T}}\bullet \nabla n)\mathbf{\mathbf{N}}-(\mathbf{\ 
\mathbf{T}}\bullet \nabla n)^{\prime }\mathbf{\mathbf{T}}\right]  \\
&& \\
&=&\left[ \left( \frac{1}{\kappa n}\right) ^{\prime }\kappa n-\frac{1}{n}(%
\mathbf{\mathbf{T}}\bullet \nabla n)\right] \mathbf{\mathbf{N}}-\frac{1}{%
\kappa n}(\mathbf{\mathbf{T}}\bullet \nabla n)^{\prime }\mathbf{\mathbf{T}}+%
\frac{1}{\kappa n}\left( \nabla n\right) ^{\prime }.
\end{eqnarray*}

\noindent By Theorem \ref{thm5}, the torsion of a light ray is determined by
the component of $(\nabla n)^{\prime }$ in the direction of $\mathbf{\mathbf{%
B}}$. But for a particle falling in a central force field, $n(r)=\sqrt{%
r_{0}^{n}r^{n}/\left( r_{0}^{n}-r^{n}\right) }$, where $r$ is the distance
from the origin and $r_{0}$ is the initial distance from the origin. Let $\hat{\mathbf{e}}_{\mathbf{r}}$ denote the unit radial vector. Then, $\nabla n=\frac{dn}{dr}\mathbf{\hat{e}_{r}}$ and $(\nabla n)^{\prime }=\left( 
\frac{dn}{dr}\right) ^{\prime }\mathbf{\hat{e}_{r}}$. Thus, $\left( \nabla n\right) ^{\prime }$ and $\nabla n$ are parallel and lie in the osculating plane. Consequently $\tau =0$ and Brachistochrone solutions in central force fields are planar. Futhermore, since $\nabla n$ is parallel to the line from the particle to the origin, the osculating plane passes through the origin and contains the initial and terminal point of the curve. But, this is exactly the problem
we investigated when we confined the particle to a plane. Therefore, a
solution curve with critical radius $c_{0}$ and initial point $A$ generates
a surface of revolution whose meridians are solution curves. These remarks are summarized in our concluding theorem:
\begin{theorem}
\label{final thm} If $\alpha(s)$ is a solution curve lying in a central force field with critical radius $c_0$, then $\tau$ is identically $0$ and $\alpha(s)$ is a meridian of the surface of revolution generated  by the curve (\ref{inverse2}).
\end{theorem}

\end{document}